\documentclass{article}
\usepackage{amssymb}
\usepackage[centertags]{amsmath}

\begin{document}

\title{Infinite-dimensional Hamilton-Jacobi theory and $L$-integrability }

\author{ Cheng-shi Liu \\Department of Mathematics\\Daqing Petroleum Institute\\Daqing 163318, China
\\Email: chengshiliu-68@126.com}

 \maketitle

\begin{abstract}
The classical Liouvile integrability means that there exist $n$
independent first integrals in involution for $2n$-dimensional phase
space. However, in the infinite-dimensional case,  an infinite
number of independent first integrals in involution don't indicate
that the system is solvable. How many first integrals do we need in
order to make the system solvable? To answer the question, we obtain
an infinite dimensional Hamilton-Jacobi theory, and prove an
infinite dimensional Liouville theorem. Based on the theorem, we
give a modified definition of the Liouville integrability in
infinite dimension. We call it the $L$-integrability. As examples,
we prove that the string vibration equation and the KdV equation are
$L$-integrable. In general, we show that an infinite number of
integrals is complete if  all action variables of a Hamilton system
can reconstructed by the set of first integrals.


\end{abstract}

\section{Introduction}
It is well-known that an infinite dimensional Hamilton system such
as KdV equation can be considered as a complete integrable
system[1,2] if it can be solved by the inverse scattering
method[3,4]. In other words, if we can obtain its all action-angle
variables, we call this system to be complete integrable and
sometimes the Liouville integrable. Finite-dimensional Liouville
theorem[5] says that if there exist $n$ independent first integrals
in evolution, the Hamilton system is solvable. The so-called
liouville integrability is just based on the Liouville theorem. For
a given Hamilton system in infinite dimension, a necessary condition
in order to make such system integrable is that it posses an
infinite number of constants of motion (or called first integrals).
F. Calogero[6] point out that due to the ambiguities in the counting
of infinities, this condition is not sufficient. A natural problem
is how many constants of motion are sufficient to ensure that such
system is solvable. We call this problem the Calogero's problem.
Indeed, we are short of an infinite-dimensional Liouville theorem
from which the above question can be solved naturally.

On the other hand, all action variables gives "all" constants of
motion. Therefore, beginning from the action variables, we can
understand the essence of the Calogero's problem on infinite number
of constants of motion. But classical Liouville integrability is not
equivalent to the solvability of the action-angle variables in
infinite dimension.  In order to give an equivalent definition, we
need to modify the conditions in the classical Liouville
integrability such that the new Liouville integrability is
equivalent to the solvability of the action-angle variables. At the
same time, the Calogero's problem is solved naturally. This is a way
from back to head. But we need a way from head to back!

In the present paper, our aim is to extend the Hamilton-Jacobi
theory and Liouville integrability (for example, see Arnold's book
[5]) from the finite-dimensional case to the infinite-dimensional
case. We establish an infinite-dimensional Hamilton-Jacobi theory,
and prove an infinite-dimensional Liouville theorem. Furthermore,
based on the theorem, a modified definition of the Liouville
integrability in infinite dimension is given. We call it
$L$-integrability. As the application of the theory, we study the
string vibration problem in detail. We  use the Hamilton-Jacobi
theory to solve it. We give an infinite number of first integrals
and prove that this is a complete set, that is, the string vibration
problem is $L$-integrable. We also discuss the problem about the
uncomplete first integrals. Finally, as an important example, we
prove that the KdV equation is $L$-integrable. Our results answer
the Calogero's problem.

We must point out that the advantage of  our theory is not in
technique but in concept since it is difficult to solve
Hamilton-Jacobi equation directly by the method of the variables
separation.

Other definitions of integrability such as Lax integrability,
$C$-integrability, can be found in Ref.[4].

\section{Infinite-dimensional Hamilton-Jacobi theory }

We consider the case of countably infinite variables.
$P=(p_1,\cdots,p_n,\cdots)$ and $Q=(q_1,\cdots,q_n,\cdots)$ are a
pair of canonical variables. $H=H(P,Q,t)$ is the Hamilton function.
The Hamilton canonical equations are as follows
\begin{equation}
\frac{\mathrm{d}q_i}{\mathrm{d}t}=\frac{\partial H}{\partial p_i},
\end{equation}
\begin{equation}
\frac{\mathrm{d}p_i}{\mathrm{d}t}=-\frac{\partial H}{\partial q_i},
\end{equation}
for $i=1,2,\cdots.$  $S$ denote the action function which takes its
value on the classical path. We have $p_i=\frac{\partial S}{\partial
q_i}$ for $i=1,2,\cdots$ and denote them by $P=\frac{\partial
S}{\partial Q}$ for simplicity. The Hamilton-Jacobi equation is
given by
\begin{equation}
\frac{\partial S}{\partial t}=-H(Q, \frac{\partial S}{\partial
Q},t).
\end{equation}
If we have a general integral $S=S(Q,\alpha)$ of the H-J equation,
where $\alpha=(\alpha_1,\alpha_2,\cdots)$, we can solve the Hamilton
canonical equation. A crucial step is that we must can solve out
$Q=Q(t,\alpha, \beta)$ from the following system of equations
\begin{equation}
\frac{\partial S}{\partial \alpha_i}=\beta_i,
\end{equation}
for $i=1,2,\cdots $, where $\beta=(\beta_1,\beta_2,\cdots)$. In the
finite dimensional case, this condition can be represented as
$\det\frac{\partial^2 S}{\partial q_i\alpha_j}\neq0$.  In the
infinite dimensional case, we use the invertible property of the
operator $\frac{\partial^2 S}{\partial q_i\alpha_j}$ instead of
$\det\frac{\partial^2S}{\partial q_i\alpha_j}\neq0$. It is easy to
prove the following result.

\textbf{Theorem 1}. If the operator $\frac{\partial^2 S}{\partial
q_i\alpha_j}$ is invertible,
\begin{equation}
Q=Q(t,\alpha, \beta),
\end{equation}
\begin{equation}
P=P(t,\alpha, \beta),
\end{equation}
are the solutions of the Hamilton canonical equations (1) and (2).

\textbf{Proof}. From $ \frac{\mathrm{d}}{\mathrm{d}t}\frac{\partial
S}{\partial \alpha_i}= \frac{\mathrm{d}\beta_i}{\mathrm{d}t}=0$, we
have
\begin{equation}
\frac{\partial^2 S}{\partial \alpha_i\partial
t}+\sum_{j=1}^{+\infty}\frac{\partial^2 S}{\partial \alpha_i\partial
q_j}\frac{\mathrm{d}q_j}{\mathrm{d}t}=0.
\end{equation}
From the Hamilton-Jacobi equation, we have
\begin{equation}
\frac{\partial^2 S}{\partial \alpha_i\partial
t}+\sum_{j=1}^{+\infty}\frac{\partial H}{\partial p_j}\frac{\partial
q_j}{\partial \alpha_i}+\sum_{j=1}^{+\infty}\frac{\partial
^2S}{\partial \alpha_i\partial q_j}\frac{\partial q_j}{\partial
\alpha_i}+\sum_{j=1}^{+\infty}\frac{\partial H}{\partial
q_j}\frac{\partial p_j}{\partial \alpha_i}=0,
\end{equation}
that is,
\begin{equation}
\frac{\partial^2 S}{\partial \alpha_i\partial
t}+\sum_{j=1}^{+\infty}\frac{\partial H}{\partial
p_j}\frac{\partial^2S}{\partial \alpha_i\partial
q_j}+\sum_{j=1}^{+\infty}\frac{\partial q_j}{\partial
\alpha_i}\frac{\partial }{\partial q_j}(\frac{\partial S}{\partial
t}+H)=\frac{\partial^2 S}{\partial \alpha_i\partial
t}+\sum_{j=1}^{+\infty}\frac{\partial H}{\partial
p_j}\frac{\partial^2S}{\partial \alpha_i\partial q_j}=0.
\end{equation}
Comparing (7) with (9), and using the condition that the operator
$\frac{\partial^2 S}{\partial q_i\alpha_j}$ is invertible, we obtain
$\frac{\mathrm{d}q_i}{\mathrm{d}t}=\frac{\partial H}{\partial p_i}$.
It is just Eq.(1).

From $p_i=\frac{\partial S}{\partial q_i}$, we have
\begin{equation}
\frac{\mathrm{d}p_i}{\mathrm{d}t}=\frac{\partial^2 S}{\partial
q_i\partial t}+\sum_{j=1}^{+\infty}\frac{\partial^2 S}{\partial
q_j\partial q_i}\frac{\mathrm{d}q_j}{\mathrm{d}t}=\frac{\partial^2
S}{\partial q_i\partial t}+\sum_{j=1}^{+\infty}\frac{\partial^2
S}{\partial q_j\partial q_i}\frac{\partial H}{\partial p_j}.
\end{equation}
By H-J equation, we have
\begin{equation}
\frac{\partial^2 S}{\partial q_i\partial t}+\frac{\partial
H}{\partial q_i}+\sum_{j=1}^{+\infty}\frac{\partial H}{\partial
p_j}\frac{\partial p_j}{\partial q_i}=0,
\end{equation}
that is
\begin{equation}
\frac{\partial^2 S}{\partial q_i\partial t}+\frac{\partial
H}{\partial q_i}+\sum_{j=1}^{+\infty}\frac{\partial^2S}{\partial
q_i\partial q_j}\frac{\partial H}{\partial p_j}=0.
\end{equation}
Comparing (10) with (12), we obtain $
\frac{\mathrm{d}p_i}{\mathrm{d}t}=-\frac{\partial H}{\partial q_i}$.
It is just Eq.(2). So we complete the proof.

\section{Infinite-dimensional Liouville theorem and $L$-integrability}

We first generalize the Liouville theorem to the infinite
dimension.\\

\textbf{Theorem 2}. Suppose that the Hamilton system has an infinite
number of first integrals (or motion constants)
\begin{equation}
f_i(P,Q,t)=\alpha_i, \ \ i=1,2,\cdots.
\end{equation}
If these first integrals satisfy the following conditions,  the
Hamilton system is integrable.

$1^0$. $[f_i, f_j]=0$, where
$[f_i,f_j]=\sum_{k=1}^{+\infty}(\frac{\partial f_i}{\partial
q_k}\frac{\partial f_j}{\partial p_k}-\frac{\partial f_i}{\partial
p_k}\frac{\partial f_j}{\partial q_k})$ is the Poisson bracket.

$2^0$. The operator $(\frac{\partial f_i}{\partial p_j})$ is
invertible, where $(\frac{\partial f_i}{\partial p_j})$ denotes the
infinite-dimensional matrix with the general element $\frac{\partial
f_i}{\partial p_j}$.

\textbf{Proof}. According to the condition $2^0$, we can solve out
\begin{equation}
p_i=\phi_i(Q,\alpha,t), \ \ i=1,2,\cdots,
\end{equation}
from the system of equations (13).  If there exists a function
$S=S(Q,\alpha,t)$ such that
\begin{equation}
\mathrm{d}S=\sum_{i=1}^{+\infty} p_i\mathrm{d}q_i-H^{*}\mathrm{d}t,
\end{equation}
that is,
\begin{equation}
\frac{\partial S}{\partial q_i}=p_i=\phi_i, \ \ i=1,2,\cdots,
\end{equation}
\begin{equation}
\frac{\partial S}{\partial t}=-H^*,
\end{equation}
where
\begin{equation}
H^*(Q,\alpha,t)=H(Q,\phi,t),
\end{equation}
according to theorem 1, we know that the Hamilton system (1) and (2)
is solvable. Indeed, the crucial step of using H-J equation method
is to solve out $\alpha$ from the system of equations
$p_i=\phi_i(Q,\alpha,t)$, for $i=1,2,\cdots$. Since $\alpha$ has
been given by Eq.(13), Hamilton system is solvable according to
theorem 1. Now what we need is to prove that the differential form
$\sum p_i\mathrm{d}q_i-H^{*}\mathrm{d}t$ is an exact form. This is
equivalent to the following conditions
\begin{equation}
\frac{\partial \phi_i}{\partial q_k}=\frac{\partial \phi_k}{\partial
q_i}, \ \ i,k=1,2,\cdots,
\end{equation}
\begin{equation}
\frac{\partial \phi_i}{\partial t}=-\frac{\partial H^*}{\partial
q_i}, \ \ i=1,2,\cdots.
\end{equation}

We first prove condition (19). Differentiating the expression (13)
with respect to $q_i$ yields
\begin{equation}
\frac{\partial f_r}{\partial q_i}+\sum_{j=1}^{+\infty}
\frac{\partial f_r}{\partial \phi_j}\frac{\partial \phi_j}{\partial
q_i}=0.
\end{equation}
We multiply  $\frac{\partial f_s}{\partial p_i}$ in two sides of
Eq.(21) and take summation for $i$. Then we have
\begin{equation}
\sum_{i=1}^{+\infty}\frac{\partial f_r}{\partial q_i}\frac{\partial
f_s}{\partial p_i}+\sum_{i,j=1}^{+\infty} \frac{\partial
f_s}{\partial p_i}\frac{\partial f_r}{\partial \phi_j}\frac{\partial
\phi_j}{\partial q_i}=0.
\end{equation}
By the same method, we have
\begin{equation}
\sum_{j=1}^{+\infty}\frac{\partial f_r}{\partial p_j}\frac{\partial
f_s}{\partial q_j}+\sum_{i,j=1}^{+\infty} \frac{\partial
f_r}{\partial p_j}\frac{\partial f_s}{\partial \phi_i}\frac{\partial
\phi_i}{\partial q_j}=0.
\end{equation}
By subtraction of  Eq.(22) and Eq.(23), and usage of $[f_r,f_s]=0$,
we have
\begin{equation}
\sum_{i,j=1}^{+\infty}\frac{\partial f_s}{\partial
q_i}\frac{\partial f_r}{\partial p_j}(\frac{\partial
\phi_j}{\partial p_i}-\frac{\partial \phi_i}{\partial p_j})=0.
\end{equation}
Since the operator $(\frac{\partial f_s}{\partial p_i})$ is
invertible by condition $2^0$, we have
\begin{equation}
\sum_{j=1}^{+\infty}\frac{\partial f_r}{\partial p_j}(\frac{\partial
\phi_j}{\partial p_i}-\frac{\partial \phi_i}{\partial p_j})=0.
\end{equation}
By the same method,  we have
\begin{equation}
\frac{\partial \phi_j}{\partial p_i}=\frac{\partial \phi_i}{\partial
p_j}.
\end{equation}

We next prove the condition (20). From the Hamilton canonical
equation, we have
\begin{eqnarray}
-\frac{\partial H}{\partial
q_i}=\frac{\mathrm{d}q_i}{\mathrm{d}t}=\frac{\mathrm{d}\phi_i}{\mathrm{d}t}=\frac{\partial
\phi_i}{\partial t}+\sum_{j=1}^{+\infty}\frac{\partial
\phi_i}{\partial
q_j}\frac{\mathrm{d}q_j}{\mathrm{d}t}=\frac{\partial
\phi_i}{\partial t}+\sum_{j=1}^{+\infty}\frac{\partial
\phi_i}{\partial q_j}\frac{\partial H}{\partial p_j},
\end{eqnarray}
so we have
\begin{eqnarray}
\frac{\partial \phi_i}{\partial t}=-\frac{\partial H}{\partial
q_i}-\sum_{j=1}^{+\infty}\frac{\partial \phi_i}{\partial
q_j}\frac{\partial H}{\partial p_j}=-\frac{\partial H}{\partial
q_i}-\sum_{j=1}^{+\infty}\frac{\partial \phi_j}{\partial
q_i}\frac{\partial H}{\partial p_j}=-\frac{\partial H^*}{\partial
q_i}.
\end{eqnarray}
We complete the proof.

Based on the above theorem, we give the following definitions.

\textbf{Definition 1}. An infinite number of motion constants (or
first integrals)(13) is called a complete set of   motion constants
if the condition $2^0$ is satisfied.

\textbf{Definition 2}. If a Hamilton system has a complete set of
motion constants,  the system is called to possess the $L$-
integrability or to be $L$-integrable.\\

\textbf{Remark 1}. In $2n$-dimensional phase space case, Liouville
integrability needs $n$ independent first integrals in involution,
which are not sufficient in infinite dimension since we can take
away some first integrals, for example, arbitrary finite number of
first integrals, such that two conditions of involution and
independence are remained as before. In our new definition 2, we
take the completeness instead of involution and independence.
Theorem 2 is the theoretical foundation of definition 2. If a
soliton equation posses an infinite number of independent first
integrals in involution,  what we need is only to verify whether
this set of first integrals is complete. We will take some concrete
examples as verification.

\section{The $L$-integrability of an infinite-dimensional harmonic oscillator}

Consider the string vibration equation
\begin{equation}
u_{tt}=u_{xx},
\end{equation}
\begin{equation}
u(0,t)=u(2\pi,t)=0,
\end{equation}
\begin{equation}
u(x,0)=u_0(x), \ \ u_t(x,0)=u_1(x),
\end{equation}
which is an infinite-dimensional harmonic oscillator. The
corresponding Lagrangian function and Hamilton function are
\begin{equation}
L=\frac{1}{2}\int_0^{2\pi} ((u_t)^2-(u_x)^2)\mathrm{d}x,
\end{equation}
and
\begin{equation}
H=\frac{1}{2}\int_0^{2\pi} ((u_t)^2+(u_x)^2)\mathrm{d}x.
\end{equation}
Let $q=u$ and $p=u_t$ be a pair of canonical variables. Then
Hamilton function is rewritten as
\begin{equation}
H(p,q)=\frac{1}{2}\int_0^{2\pi} (p^2+(q_x)^2)\mathrm{d}x.
\end{equation}
Therefore the Hamilton canonical system is just Eq.(29). The
Hamilton-Jacobi equation is given by
\begin{equation}
\frac{\partial S}{\partial t}=-\frac{1}{2}\int_0^{2\pi}
\{(\frac{\delta S}{\delta q})^2+(q_x)^2\}\mathrm{d}x
\end{equation}
By the separation of variables, we let
\begin{equation}
S(q,t)=S_0(t)+S_1(q),
\end{equation}
where $S_1(q)$ is a functional of $q$. Substituting Eq.(36) into
Hamilton-Jacobi equation (35) and separating the variables yield
\begin{equation}
S'_0(t)=-\frac{1}{2}\int_0^{2\pi} \{(\frac{\delta S_1}{\delta
q})^2+(q_x)^2\}\mathrm{d}x=-\frac{1}{2}\int_0^{2\pi}\beta(x)\mathrm{d}x=-E.
\end{equation}
It follows that
\begin{equation}
S_0(t)=-Et,
\end{equation}
\begin{equation}
(\frac{\delta S_1}{\delta q})^2+(q_x)^2=\beta(x),
\end{equation}
where $\beta(x)$ is an arbitrary function satisfying
$\frac{1}{2}\int_0^{2\pi}\beta(x)\mathrm{d}x=E$. For example, we can
take $\beta(x)=2E$. But we must point out that Eq.(39) is a
variation-differential equation and we don't know how to solve it in
general case.

Now we adopt another method to deal with these problems. Take
Fourier transformation of $u$ with respect to $x$,
\begin{equation}
u(x,t)=\sum_{n=1}^{+\infty}a_n(t)\sin (nx),
\end{equation}
then
\begin{equation}
u_t(x,t)=\sum_{n=1}^{+\infty}a'_n(t)\sin (nx).
\end{equation}
Therefore the Hamiltonian function becomes
\begin{equation}
H=\frac{1}{2}\sum_{n=1}^{+\infty}\{a'^2_n(t)+n^2a^2_n(t)\}.
\end{equation}
Taking
\begin{equation}
S(a_1(t), a_2(t),\cdots)=S_0(t)+\sum_{n=1}^{+\infty}S_n(a_n),
\end{equation}
and substituting it into Hamilton-Jacobi equation yield
\begin{equation}
(\frac{\mathrm{d}S_n}{\mathrm{d}a_n})^2+n^2a^2_n=E_n, \ \
n=1,2,\cdots,
\end{equation}
where $E_n's$ are constants and satisfy the following condition
\begin{equation}
\sum_{n=1}^{+\infty}E_n=2E.
\end{equation}
Solving Eq.(44), we have
\begin{equation}
S_n=\int\sqrt{E_n-n^2a^2_n}\mathrm{d}a_n.
\end{equation}
According to the standard steps we can solve out the solutions of
$a_n$, for $n=1,2,\cdots$. Hence we can use the Hamilton-Jacobi
theory to solve the string vibration problem.

We next obtain the $L$-integrability of the string vibration
equation. We first give an infinite number of first integrals
\begin{equation}
f_n(u,u_t)=\frac{1}{2}n^2(\int_0^{2\pi}u(x,t)\sin(nx)\mathrm{d}x)^2+
\frac{1}{2}(\int_0^{2\pi}u_t(x,t)\sin(nx)\mathrm{d}x)^2,
\end{equation}
for $n=1,2,\cdots$.  Indeed, we have
\begin{eqnarray}
\frac{\mathrm{d}}{\mathrm{d}t}f_n(u,u_t)=n^2\int_0^{2\pi}u(x,t)\sin(nx)\mathrm{d}x
\int_0^{2\pi}u_t(x,t)\sin(nx)\mathrm{d}x\cr+
\int_0^{2\pi}u_t(x,t)\sin(nx)\mathrm{d}x\int_0^{2\pi}u_{tt}(x,t)\sin(nx)\mathrm{d}x=0,
\end{eqnarray}
where we use $u_{tt}=u_{xx}$ and  integration by part in last step.
Rewriting the first integrals in terms of variables $a_n's$, we have
\begin{equation}
f_n=\frac{1}{2}({a'}_n^2(t)+n^2a_n^2(t)),\ \ n=1,2,\cdots.
\end{equation}
Therefore, every $f_n$ is just the energy of the $n$th mode. The
physical picture of these first integrals is very clear.

We now prove these first integrals constitute a complete set.
Indeed, in this case, the canonical variables are $q_n=a_n$ and
$p_n=a'_n$. From the set of first integrals (49) represented by
$f_n=\frac{1}{2}(p_n^2(t)+n^2q_n^2(t))$  in terms of $q_n$ and
$p_n$, we can solve out the $p_n's$. It follows that this is a
complete set. On the other hand, we have
\begin{equation}
\frac{\partial f_n}{\partial q_m}=\delta mn,
\end{equation}
where $\delta mn$ is the Dirac sign in infinite dimension, that is,
 the operator (matrix) $(\frac{\partial f_n}{\partial q_m})$ is
invertible. Of course, $[f_n,f_m]=0$ is a simple fact. According to
theorem 2, the string vibration problem is  $L$-integrable.

If we remove some first integrals in the set (49),  the set will be
not complete. Indeed, for example, we remove $f_1$, then the remains
are also evolutional and independent. But it is easy to see that the
remains are not complete since we can't solve out $p_1$. In other
words, the operator (matrix) $(\frac{\partial f_n}{\partial
q_m})_{m=1,n=2}^{+\infty}$ isn't invertible.

\section{The $L$-integrability of the infinite vibrating string}

We consider the Cauchy problem for an infinite vibrating string
\begin{equation}
u_{tt}=u_{xx},
\end{equation}
\begin{equation}
u(-\infty,t)=u(+\infty,t)=0,
\end{equation}
\begin{equation}
u(x,0)=u_0(x), \ \ u_t(x,0)=u_1(x).
\end{equation}
We take the Fourier transformation of $u(x,t)$ with respect to the
variable $x$,
\begin{equation}
u(x,t)=\frac{1}{2\pi}\int_{-\infty}^{+\infty}a(y,t)\sin(xy)\mathrm{d}y.
\end{equation}
It is easy to prove that
\begin{equation}
f(y,t)=\frac{1}{2}\{{a_t}^2(y,t)+y^2a^2(y,t)\},
\end{equation}
or in another form
\begin{equation}
f(y,t)=\frac{1}{2}\{(\frac{1}{2\pi}\int_{-\infty}^{+\infty}u_t(x,t)\sin(xy)\mathrm{d}x)^2
+y^2(\frac{1}{2\pi}\int_{-\infty}^{+\infty}u(x,t)\sin(xy)\mathrm{d}x)^2\},
\end{equation}
is a first integral for every $y$, that is,
$\frac{\mathrm{d}}{\mathrm{d}t}f(y,t)=0$. This first integral is
just the energy of the $y$-th mode. They constitute a set of the
first integrals with uncountably infinite elements. In order to
construct a countably infinite number of first integrals, we take
the Taylor expansion of $f(y,t)$ with respect to the variable $y$
\begin{equation}
f(y,t)=\sum_{n=o}^{+\infty}\frac{\mathrm{d^n}f(0,t)}{\mathrm{d}t^n}y^n.
\end{equation}
Then every $\frac{\mathrm{d^n}f(0,t)}{\mathrm{d}t^n}$ is a first
integral. Through tedious computation, we obtain
\begin{equation}
f(y,t)=g_1y^2+g_2y^4+\cdots+g_ky^{2k}+\cdots,
\end{equation}
where
\begin{equation}
g_1=(\int_{-\infty}^{+\infty} xu_t(x,t)\mathrm{d}x)^2,
\end{equation}
\begin{eqnarray}
g_k=\frac{(-1)^{k+1}}{(2k-1)!}\int_{-\infty}^{+\infty}
xu_t(x,t)\mathrm{d}x\int_{-\infty}^{+\infty}
x^{2k-1}u_t(x,t)\mathrm{d}x\cr
+\sum_{m=0}^{k-2}\frac{(-1)^{k}}{(2m+1)!(2(k-m)-3)!}\{\int_{-\infty}^{+\infty}
x^{2(k-m)-3}u\mathrm{d}x\int_{-\infty}^{+\infty}
x^{2m+1}u\mathrm{d}x\cr
-\frac{1}{(2(k-m)-2)(2(k-m)-1)}\int_{-\infty}^{+\infty}
x^{2(k-m)-1}u_t\mathrm{d}x\int_{-\infty}^{+\infty}
x^{2m+1}u_t\mathrm{d}x\},
\end{eqnarray}
for $k=2,3,\cdots.$ Every $g_n$ is a first integral. We will prove
that they constitute a complete set of first integrals. For the
purpose, we first take the canonical coordinates as
\begin{equation}
q_n=\int_{-\infty}^{+\infty} x^{2n+1}u\mathrm{d}x,
\end{equation}
\begin{equation}
p_n=\int_{-\infty}^{+\infty} x^{2n+1}u_t\mathrm{d}x,
\end{equation}
for $n=0,1,\cdots.$ In terms of the canonical coordinates, we
rewrite $g_k$ as
\begin{equation}
g_1=p_0^2,
\end{equation}
\begin{eqnarray}
g_k=\frac{(-1)^{k+1}}{(2k-1)!} p_0 p_{k-1}
+\sum_{m=0}^{k-2}\frac{(-1)^{k}}{(2m+1)!(2(k-m)-3)!}\cr\times\{
q_{k-m-2} q_{m}-\frac{1}{(2(k-m)-2)(2(k-m)-1)} p_{k-m-1} p_{m}\},
\end{eqnarray}
for $k=2,3,\cdots.$ From the system of equations $g_k=\beta_k$ for
$k=0,1,\cdots$, we can easily solve out the $p_n$ for
$n=0,1,\cdots$, since these equations all are quadratic. Indeed, we
have
\begin{equation}
p_0=\pm\sqrt{\beta_1},
\end{equation}
\begin{eqnarray}
p_k=\frac{(2k+1)!}{2}\{(-1)^k\beta_{k+1}+\frac{1}{(2k-1)!} q_0
q_{k-1}-\sum_{m=1}^{k-2}\frac{(-1)^{k}}{(2m+1)!(2(k-m)-3)!}\cr\times(
q_{k-m-2} q_{m}-\frac{1}{(2(k-m)-2)(2(k-m)-1)} p_{k-m-1} p_{m})\},
\end{eqnarray}
for $k=2,3,\cdots.$ Thus we conclude that the infinite vibrating
string problem is  the $L$-integrable.

We must notice that the expression of $g_k$ is so complicate  that
we can't clearly find the physical meanings of these first
integrals. On the other hand, the physical meaning of the first
integral $f(y,t)$ is very clear.  Thus, behind those complicate
first integrals, perhaps there is a simple rule such that a clear
physical picture can be emerged to us.

\textbf{Remark 2}. It is easy to see that
\begin{equation}
g_n(t)=\int_{-\infty}^{+\infty} x^n u_t(x,t)\mathrm{d}x
\end{equation}
are first integrals for $n=0,1,\cdots$ and they constitute a
countably infinite set of first integrals. But this is not a
complete set of first integrals.

\section{The $L$-integrability of the KdV equation}

Consider the following KdV equation
\begin{equation}
u_t-6uu_x+u_{xxx}=0.
\end{equation}
The corresponding Schrodinger equation is
\begin{equation}
-\frac{\mathrm{d}^2}{\mathrm{d}x^2}\phi+u\phi=k^2\phi.
\end{equation}
As $x\longrightarrow\infty$ for $\mathrm{Im}k>0$ and  $u$ satisfies
the KdV equation,  we let
 \begin{equation}
a(k)=\phi(x,k)\mathrm{e}^{i kx},
\end{equation}
 we have (see Ref.[3] for the details on inverse
scattering transformation)
\begin{equation}
\frac{\mathrm{d}}{\mathrm{d}t}a(k)=0.
\end{equation}
 Let
\begin{equation}
\ln a(k)=\int_{-\infty}^{+\infty}\chi(x,k)\mathrm{d}x.
\end{equation}
Then $\chi$ satisfies the Riccati's equation
\begin{equation}
\chi_x+\chi^2-u-2ik\chi=0.
\end{equation}
Furthermore, let
\begin{equation}
\chi=\sum_{m=1}^{+\infty}\frac{\chi_m}{(2ik)^m},
\end{equation}
we have $\int\chi_{2m}\mathrm{d}x=0$ and an infinite number of first
integrals
\begin{equation}
I_m=\int\chi_{2m-1}\mathrm{d}x=0,
\end{equation}
for $m=0,1,\cdots.$ Now we can prove that these first integrals
constitute a complete set since we can determine the function $a(k)$
by the values of these first integrals $I_m$ for $m=0,1,\cdots$.
Indeed, if we know the function $a(k)$, we can obtain the action
variables $n(k)=\frac{2k}{\pi}\ln|a(k)|^2, k>0$ and $N_l=k_l^2,
k=1,\cdots,N$, where $ik_l's$ are the zeros of $a(k)$. In other
words, we can solve out the action variables. We conclude that these
first integrals constitute a complete set, that is, the KdV equation
is the $L$-integrable. Of course, we can obtain the action-angle
variables by solving the Hamilton-Jacobi equation with the
Hamiltonian rewritten in terms of the action variables
\begin{equation}
H=-\frac{32}{5}\sum_{l=1}^N
N_l^{5/2}+8\int_{0}^{+\infty}k^3n(k)\mathrm{d}k.
\end{equation}

\textbf{Remark 3}. The $L$-integrability of other soliton equations
such as nonlinear Schrodinger equation and Sine-Gordon equation, can
 be established easily.

\textbf{Remark 4}. Magri[7] use a Bi-Hamiltonian structures and the
infinitesimal symmetry transformation method  to study  an infinite
number of constants of motion.  Wadati[8] also use the infinitesimal
symmetry transformation method to obtain many results about
conversation laws of KdV equation. The modern development of
Bi-Hamiltonian structures method can be seen Ref.[9] and the
references therein.

Of course, if an infinite-dimensional Hamilton system can be
reformulated in terms of action-angle variables, we usually consider
it integrable. Here, we strictly prove that this integrability is
just the $L$-integrability. In other words, an infinite number of
first integrals represented in terms of all action variables is a
complete set. By theorem 2, we can easily prove the following
theorem.

\textbf{Theorem 3}. For a given Hamilton system, if its all action
variables can be reconstructed by an infinite number of first
integrals, the set of first integrals is complete, that is, the
system is $L$-integrable.

\section{Conclusion}
We extend the Hamilton-Jacobi theory and Liouville theorem
 from the finite-dimensional case to the
infinite-dimensional case. We introduce the $L$-integrability which
can be considered as a suitable definition of Liouville
integrability in infinite dimension. As examples, we study the
string vibration problem in detail, and use the Hamilton-Jacobi
theory to solve it. We give an infinite number of first integrals
and prove that this is a complete set, that is, the string vibration
problem is $L$-integrable. We also discuss the problem about the
uncomplete first integrals. From our discussion, the physical
picture of the $L$-integrability of the string vibration problem is
very clear. Finally, we apply our theory to the soliton equation and
prove that the KdV equation is $L$-integrable. Of course, the
$L$-integrability of other nonlinear evolution equations such as
nonlinear Schrodinger equation can also be easily obtained.\\

\textbf{Acknowledgments}. I would like to thank professor Calogero
for his valuable suggestions.

\end{document}